\documentstyle[aps,floats,prl]{revtex}
\begin{document}

\newcommand{\be}{\begin{equation}}
\newcommand{\ee}{\end{equation}}
\newcommand{\beqa}{\begin{eqnarray}}
\newcommand{\eeqa}{\end{eqnarray}}
\newcommand{\ba}{\begin{array}}
\newcommand{\ea}{\end{array}}

\title{
ULTRA HIGH ENERGY COSMIC RAYS FROM 
DECAYING SUPERHEAVY PARTICLES~\footnote{talk given at the workshop 
"Observing the Highest Energy Particles from Space", University of Maryland
, 13 - 15 November, 1997}}
      
\author{ V.Berezinsky}

\address{INFN, Laboratori Nazionali del Gran Sasso, 67010 Assergi, Italy}

\maketitle

\begin{abstract}

Decaying superheavy particles can be produced by Topological Defects or,
in case they are quasi-stable, as relics from the early Universe. The 
decays of these 
particles can be the sources of observed Ultra High Energy Cosmic Rays 
($E \sim 10^{10}\, - \, 10^{12}~GeV$). The Topological Defects as the 
UHE CR sources are critically reviewed and cosmic necklaces and 
monopole-antiminopole pairs are identified as most plausible sources.
The relic superheavy particles are shown to be clustering in the halo
and their decays produce UHE CR without GZK cutoff. The 
Lightest Supersymmetric Particles with Ultra High Energies are naturally 
produced in the cascades accompanying the decays of superheavy particles. 
These particles are discussed as UHE carriers in the Universe.    

\end{abstract}

\section{Introduction}
The observation of cosmic ray particles with energies higher than $10^{11}~GeV$
\cite{EHE}  gives a serious challenge to the known mechanisms of acceleration. 
The shock acceleration in different 
astrophysical objects typically gives maximal energy of accelerated protons
less than $(1-3)\cdot 10^{10}~GeV$ \cite{NMA}. The unipolar induction can 
provide the maximal energy $1\cdot 10^{11}~GeV$ only for the extreme values 
of the parameters \cite{BBDGP}. Much attention has recently been given to 
acceleration by ultrarelativistic shocks \cite{Vie},\cite{Wax}. The
particles here can gain a tremendous increase in energy,
equal to $\Gamma^2$, at a single reflection, 
where $\Gamma$ is the Lorentz factor of the shock.
However, it is 
 known (see e.g.  the simulation 
for pulsar relativistic wind in \cite{Hosh}) that particles entering 
the shock region are captured there or at least have a small probability 
to escape. 

{\em Topological defects, TD,} (for a review see \cite{Book}) can naturally 
produce particles of ultrahigh energies (UHE). The pioneering observation 
of this possibility was made by Hill, Schramm and Walker \cite{HS} (for 
a general analysis of TD as UHE CR sources see \cite {BHSS} and for a 
review \cite{Sigl}).

In many cases TD become unstable and decompose to constituent fields, 
superheavy gauge and Higgs bosons (X-particles), which then decay 
producing UHE CR. It could happen, for example, when two segments of 
ordinary string, or monopole and antimonopole touch each other, when 
electrical current in superconducting string reaches the critical value
and in some other cases.

In most cases the problem with UHE CR from TD 
is not the maximal energy, but the fluxes. One very general reason 
for the low fluxes consists in the large distance between TD. A dimension
scale for this distance is the Hubble distance $H_0^{-1}$. However, in some 
rather exceptional cases this dimensional scale is multiplied to a small 
dimensionless value $r$. If a distance between TD is larger than 
 UHE proton attenuation length (due to the GZK effect \cite{GZK}), then 
the flux at UHE is typically exponential suppressed.

{\em Ordinary cosmic strings} can produce particles when a loop annihilate 
into double line \cite{BR}. The produced UHE CR flux is strongly reduced due 
to the fact that a loop oscillates, and in the process of a collapse the 
two incoming parts of a loop touch each other in one point producing thus 
the smaller loops, instead of two-line annihilation. However, this idea was 
recently revived due to recent work \cite{Vincent}. It is argued there that 
the energy loss of the long strings is dominated by production of very 
small loops with the size down to the width of a string, which immediately 
annihilate into superheavy particles. A problem with this scenario is 
too large distance between strings (of order of the Hubble distance). 
For a distance between an observer and a string being the same, the 
observed spectrum of UHE CR has an exponential cutoff at energy 
$E \sim 3\cdot 10^{19}~eV$.

Superheavy particles can be also produced when two segments of
string come 
into close contact, as in {\it cusp} events \cite{Bran}.  
This process
was studied later by Gill and Kibble \cite{GK}, and they concluded that
the resulting cosmic ray flux is far too small.  
An interesting possibility suggested by Brandenberger  \cite{Bran} is the
{\em cusp} ``evaporation'' on cosmic strings. 
When the distance between two segments of the cusp
becomes of the order of the string width, the cusp may``annihilate" 
turning into high energy particles, 
which are boosted by a very large Lorentz
factor of the cusp \cite{Bran}.  
However, the resulting UHE CR flux is considerably smaller than one 
observed \cite{BBM}.  

{\em Superconducting strings} \cite{Witten} appear to
be much better suited for particle production.
Moving through cosmic magnetic fields, such strings develop
electric currents and copiously produce charged heavy particles when the
current reaches certain critical value. 
The CR flux produced by
superconducting strings is affected by some model-dependent string
parameters and by the history and spatial distribution of cosmic
magnetic fields.  
Models considered so far failed to account
for the observed flux \cite{SJSB}.

{\em Monopole-antimonopole pairs } ($M{\bar M}$) 
can form bound states and eventually
annihilate into UHE particles  \cite{Hill}, \cite{BS}.  
For an appropriate choice of the
monopole density $n_M$, this model is consistent with observations;
however, the required (low) value of $n_M$ implies fine-tuning. 
In the first phase transition $G \to H \times U(1)$ in the early 
Universe the monopoles are produced with too high density. It must then be 
diluted by inflation to very low density, precisely tuned to 
the observed UHE CR flux. 

{\em Monopole-string network} can be formed in the early Universe in the 
sequence of symmetry breaking 
\be
G \to H \times U(1) \to H \times Z_N.
\label{eq:Z-N}
\ee
For $N \geq 3$ an infinite network of monopoles connected by strings is 
formed. The magnetic fluxes of monopoles in the network are channeled into 
into the strings that connect them. The monopoles typically have additional 
unconfined magnetic and chromo-magnetic charges. When strings shrink the 
monopoles are pulled by them and are accelerated. The 
accelerated monopoles produce extremely high energy gluons, which then 
fragment into UHE hadrons \cite{BMV}. The produced flux is too small 
to explain UHE CR observation \cite{BBV}.

{\em Cosmic necklaces} are TD which are formed  
in a sequence of symmetry breaking given by Eq.(\ref{eq:Z-N}) when $N=2$. 
 The first phase transition
produces monopoles, and at the second phase transition each monopole
gets attached to two strings, with its magnetic flux channeled along the 
strings.  The resulting necklaces resemble ``ordinary'' cosmic strings
with monopoles playing the role of beads. Necklaces can evolve in such way
that a distance between monopoles diminishes and in the end all monopoles 
annihilate with the neighboring antimonopoles \cite{BV}. The produced 
UHE CR flux \cite{BV} is close to the observed one and the shape of the 
spectrum 
resembles one observed. The distance between necklaces can be much smaller 
than attenuation length of UHE protons.

{\em Superheavy relic particles} can be sources of UHE CR \cite{KR,BKV}.
In this scenario Cold Dark Matter (CDM) have a small admixture 
of long-lived superheavy particles. These particles must be heavy,
$m_X > 10^{12}~GeV$, long-lived $\tau_X > t_0$, where $t_0$ is the age 
of the Universe, and weakly interacting. The required life-time 
can be 
provided if this particle has (almost) conserved quantum number broken 
very weakly due to warmhole \cite{BKV} or instanton \cite{KR} effects.
Several mechanisms for production of such particles in the early Universe 
were identified. Like other forms of non-dissipative CDM , X-particles 
must accumulate in the halo of our Galaxy \cite{BKV} and thus they produce 
UHE CR without GZK cutoff and without appreciable anisotropy. 

{\em The UHE carriers} produced at the decay of superheavy 
relic particles or from TD, can be  be nucleons,
photons and neutrinos or neutralinos \cite{BK}. Production of neutralinos
occurs in particle cascade, which originates at the decay of superheavy 
X-particle in close analogy to QCD cascade \cite{BK}. Though flux of 
UHE neutralino is of the same order as neutrino flux, its detection is more 
problematic because of smaller cross-section. The other particles 
discussed as carrier of UHE signal are gluino \cite{Farr,MN,BK}, in case it is 
Lightest Supersymmetric Particle (LSP), and heavy monopole \cite{Weil,MN}.  

In this paper I will present the results obtained in our joint works with
M.Kachelriess and A.Vilenkin about necklaces and  superheavy relic 
particles as possible sources of UHE CR. Neutralino and gluino as the 
carriers of UHE signal will be also shortly discussed.

\section{Necklaces}  
Necklaces produced in a sequence of symmetry breaking 
$G \to H \times U(1) \to H \times Z_2$ form the infinite necklaces having 
the shape or random walks and a distribution of closed loops. Each monopole 
in a necklace is attached to two strings. 

The monopole mass $m$ and the string tension $\mu$ are determined by
the corresponding symmetry breaking scales, $\eta_s$ and $\eta_m$ 
($\eta_m>\eta_s$): 
$m\sim 4\pi\eta_m/e,\;\;\; \mu\sim 2\pi\eta_s^2$.
Here, $e$ is the gauge coupling.  The mass per unit length of string
is equal to its tension, $\mu$.  Each string attached to a monopole
pulls it with a force $F \sim \mu$ in the direction of the string.  
The monopole radius $\delta_m$ and the string thickness $\delta_s$ are
typically of the order $\delta_m\sim(e \eta_m)^{-1}$, $~~\delta_s\sim
(e \eta_s)^{-1}$.  

An important quantity for the necklace evolution  is the dimensionless ratio
\begin{equation}
r=m/\mu d,
\label{r}
\end{equation}

We expect the necklaces to evolve in a 
scaling regime. If $\xi$ is the characteristic length scale of the network, 
equal to the typical separation of long strings and to their characteristic
curvature radius, then the force  per unit length of string is 
$f \sim \mu/\xi$, and the acceleration is $a \sim  
(r+1)^{-1}\xi^{-1}$.
We assume that $\xi$ changes on a Hubble time scale $\sim t$. Then the 
typical distance travelled by long strings in time $t$ should be 
$\sim \xi$, so that the strings have enough time to intercommute in a 
Hubble time. This gives $at^2 \sim\xi$, or 
\begin{equation}
\xi \sim (r+1)^{-1/2}t.
\label{xi}
\end{equation}
The typical string velocity is $v \sim (r+1)^{-1/2}$.

It is argued in Ref.(\cite{BV}) that $r(t)$ is driven towards large value
$r \gg1$. However, for $r \geq 10^6$ the characteristic velocity of the 
network falls down below the virial velocity, and the necklaces will be 
trapped by gravitational clustering of the matter. This may change 
dramatically the evolution of network. One possible interesting effect
for UHE CR can be enhancement of necklace space density within 
Local Supercluster -- a desirable effect as far  
absence of the GZK cutoff is concerned. However, we restrict our 
consideration by the case $r < 10^6$. The distance between 
necklaces is still small enough, $\xi \gtrsim 3~Mpc$, to assume their uniform 
distribution, when calculating the UHE CR flux.

Self-intersections of long necklaces result in copious production of
closed loops.  For $r\gtrsim 1$ the motion of loops is not periodic,
so loop self-intersections should be frequent and their fragmentation
into smaller loops very efficient.
A loop of size $\ell$ typically disintegrates on a timescale 
$\tau\sim r^{-1/2}\ell$.  All monopoles trapped in the loop must, of course, 
annihilate in the end.

Annihilating $M{\bar M}$ pairs decay into
Higgs and gauge bosons, which we shall refer to collectively as
$X$-particles.  The rate of $X$-particle production is easy to
estimate if we note that infinite necklaces lose a substantial
fraction of their length to closed loops in a Hubble time.  
The string length per unit volume is $\sim \xi^{-2}$, and the monopole
rest energy released per unit volume per unit time is
$r\mu/\xi^2 t$.  Hence, we can write
\begin{equation}
\dot{n}_X \sim r^2 \mu/(t^3 m_X),
\label{xrate}
\end{equation}
where $m_X\sim e\eta_m$ is the $X$-particle mass.

X-particles emitted by annihilating monopoles decay into hadrons, photons 
and neutrinos, the latter two components are produced through decays of 
pions.

The diffuse flux of ultra-high energy protons
can be evaluated as
\begin{equation}
I_p(E)=\frac{c\dot{n}_X}{4 \pi m_X} \int_0^{t_0} dt\,\, W_N(m_X,x_g)
\frac{dE_g(E,t)}{dE}
\label{pflux}
\end{equation}
where $dn_X/dt$ is given by Eq.(\ref{xrate}),
$E$ is an energy of proton at observation and $E_g(E,t)$ is its energy at 
generation at cosmological epoch $t$, $x_g=E_g/E$ and
$W_N(m_X,x)$ is the fragmentation function of X-particle into nucleons
of energy $E=xm_X$. The value of $dE_g/dE$ can be calculated from the 
energy losses of a proton on microwave background radiation (e.g. see 
\cite{BBDGP}). In Eq.(\ref{pflux}) the recoil protons are taken into 
account, while in Ref.(\cite{BV}) their contribution was neglected.

The fragmentation function $W_N(m_X,x)$ is calculated using the decay of 
X-particle into QCD 
partons (quark, gluons and their supersymmetric partners) with the 
consequent development of the parton cascade. The cascade in this case is
identical to one initiated by $e^+e^-$ -annihilation.
We have used  the fragmentation function in the gaussian form as 
obtained in MLLA approximation in \cite{DHMT} and
\cite{ESW}.

In our calculations 
the UHE proton flux is fully determined by only two parameters,
$r^2\mu$ and $m_X$. 
The former is restricted by low energy diffuse gamma-radiation.
It results from e-m cascades initiated 
by high energy photons and electrons produced in 
the decays of X-particles.

The cascade energy density predicted in our model is 
\begin{equation}
\omega_{cas}=\frac{1}{2}f_{\pi}r^2\mu \int_0 ^{t_0}\frac{dt}{t^3}
\frac{1}{(1+z)^4}=\frac{3}{4}f_{\pi}r^2\frac{\mu}{t_0^2},
\label{cas}
\end{equation}
where $t_0$ is the age of the Universe (here and below we use
$h=0.75$), $z$ is the redshift and $f_{\pi} \sim 1$ is the fraction of
energy 
transferred to pions. In Eq.(\ref{cas}) we took into account that half 
of the energy of pions is transferred to photons and electrons. 
The observational bound on the cascade density, for the kind of
sources we are considering here, is \cite{B92} $\omega_{cas} \lesssim
10^{-5}~eV/cm^3$.  This gives a bound on the parameter $r^2\mu$.

In numerical  calculations  we used 
$r^2\mu= 1\times 10^{28}~GeV^2$, which results in 
$\omega_{cas}=5.6 \cdot 10^{-6}~eV/cm^3$, somewhat below the observational 
limit. Now we are left with one free parameter, $m_X$, which we fix at
$1\cdot 10^{14}~GeV$.  The maximum energy
of protons is then $E_{max} \sim 10^{13}~GeV$.  
The calculated proton 
flux is presented in Fig.1, together with a summary 
of observational data taken 
from ref.\cite{akeno}. 

Let us now turn to the calculations of UHE gamma-ray flux from  the decays 
of X-particles. The dominant channel is given by the decays of neutral 
pions. The flux can be readily calculated as
\begin{equation}
I_{\gamma}(E)=\frac{1}{4\pi}\dot{n}_X\lambda_{\gamma}(E)N_{\gamma}(E)
,
\label{gflux}
\end{equation}
where $\dot{n}_X$ is given by Eq.(\ref{xrate}), 
$\lambda_{\gamma}(E)$ is the absorption length of a photon with energy 
$E$ due to $e^+e^-$ pair production on background radiation and 
$N_{\gamma}(E)$ is the number of photons with energy E produced per one decay 
of X-particle. The latter is given by 
\begin{equation}
N_{\gamma}(E)=\frac{2}{m_X}\int_{E/m_X}^1 \frac{dx}{x} W_{\pi^0}(m_X,x)
\label{gnumber}
\end{equation}
where $W_{\pi^0}(m_X,x)$ is the fragmentation function of X-particles into
$\pi^0$ pions.

At energy $E>1\cdot 10^{10}~GeV$ the 
dominant
contribution to the gamma-ray absorption comes from the radio background. The 
significance of this process was first noticed in \cite{B70}(see also 
book \cite{BBDGP}). New calculations for this absorption 
were recently
done \cite{PB}. We have used the 
absorption lengths from this work. 

When evaluating the flux \ref{gflux} at $E > 1 \cdot 10^{10}~GeV$ we neglected 
cascading of a primary photon, because pair production and inverse 
compton scattering occur at these energies on radio background, and thus 
at each collision the energy of a cascade particle is halved.
Moreover, assuming an intergalactic magnetic field 
$H \geq 1\cdot 10^{-9}$, the secondary 
electrons and positrons loose their energy mainly due to synchrotron 
radiation and the emitted photons escape from the considered energy 
interval \cite{BGP}.   

The calculated flux of gamma radiation is presented in Fig.1 by the curve
labelled $\gamma$. One can see that at $E \sim 1\cdot 10^{11}~GeV$ the 
gamma ray flux is considerably lower than that of protons. This is 
mainly due 
to the difference in the attenuation lengths for protons ($110~Mpc$) and 
photons ($2.6~Mpc$ \cite{PB} and $2.2~Mpc$ \cite{B70}). At higher energy 
the attenuation length for protons dramatically decreases ($13.4~Mpc$ at 
$E=1 \cdot 10^{12}~GeV$) and the fluxes of protons and photons become 
comparable.

A requirement for the models explaining the observed UHE events is that 
the distance between sources must be smaller than the attenuation
length. Otherwise the flux at the 
corresponding energy would be 
exponentially suppressed. This imposes a severe constraint on the
possible sources.  For example, in the case of protons with energy 
$E \sim (2- 3)\cdot 10^{11}~GeV$ 
the proton 
attenuation length is $19~Mpc$ . If protons 
propagate rectilinearly, there should be several sources inside this
radius;
otherwise all particles would arrive from the same direction.
If particles are strongly deflected in extragalactic magnetic fields, 
the distance to the source should be even smaller.  Therefore, the 
sources of 
the observed events at the highest energy must be at a distance 
$R\lesssim 15~Mpc$ in the case or protons. 

In our model the distance between 
sources, given by Eq.(\ref{xi}), satisfies this condition for 
$r>3\cdot 10^{4}$.
This is in contrast to other potential sources,
including supeconducting cosmic strings and powerful 
astronomical sources such as AGN, for which this condition imposes
severe restrictions.  

The difficulty is even more pronounced in the case of UHE photons. 
These particles 
propagate rectilinearly and their absorption length is shorter: 
$2 - 4~Mpc$ at $E \sim 3\cdot 10^{11}~GeV$. It is rather unrealistic to expect
several powerful astronomical sources at such short distances. This 
condition is very restrictive for topological defects as well. The
necklace model is rather exceptional regarding this aspect.

\section{UHE CR FROM RELIC QUASISTABLE PARTICLES}

This possibility was recognized recently in Refs(\cite{KR,BKV}).

Our main assumption is that Cold Dark Matter 
(CDM) has a small admixture of long-lived supermassive $X$-particles.
Since, apart from very small scales, fluctuations grow identically
in all components of CDM, the fraction of $X$-particles, $\xi_X$, is 
expected to be the same in all structures. In particular, $\xi_X$ is
the same in the halo of our Galaxy and in the
extragalactic space. Thus the halo density of $X$-particles is enhanced in 
comparison with the extragalactic density.
The decays 
of these particles produce UHE CR, whose flux is dominated by the 
halo component, and therefore has no GZK cutoff. Moreover,
the potentially dangerous e-m cascade radiation is suppressed.

First, we address the elementary-particle and cosmological aspects
of a superheavy long-living particle. Can the relic density 
of such particles be as high as required by observations of UHE CR?
And can they have a lifetime comparable or larger than the age
of the Universe?
 
Let us assume that $X$-particle is a neutral fermion which belongs 
to a representation of the $SU(2)\times U(1)$ group.  We assume also
that the stability of $X$-particles is protected by a discrete
symmetry 
which is the remnant of a gauge symmetry and is
respected by all interactions except quantum
gravity through wormhole effects. In other words, our particle is very
similar to a very heavy neutralino with a conserved quantum number, $R'$, 
being the direct analogue of $R$-parity (see \cite{BJV} and the
references therein).
Thus, one can assume that the decay of $X$-particle occurs due to dimension 5
operators, inversely proportional to the Planck mass $m_{\rm Pl}$ and 
additionally suppressed by a factor $\exp(-S)$, where $S$ is the 
action of a wormhole which absorbs $R'$-charge. 
As an example one can consider a term
\be
{\cal L} \sim \frac{1}{m_{Pl}} \bar{\Psi}\nu \phi\phi \exp(-S),
\label{eq:d5}
\ee
where $\Psi$ describes X-particle, and $\phi$ is a $SU(2)$ scalar with vacuum 
expectation value $v_{EW}=250$~GeV.
After spontaneous symmetry breaking the term (\ref{eq:d5}) results in 
the mixing of $X$-particle and neutrino, and  
the lifetime due to $X \to \nu +q + \bar{q}$ , {\it e.g.}, is given by
\be
\tau_X \sim \frac{192(2\pi)^3}{(G_Fv_{EW}^2)^2}\frac{m_{\rm Pl}^2}{m_X^3}
e^{2S},
\label{eq:ltime}
\ee
where $G_F$ is the Fermi constant.  The lifetime $\tau_X >t_0$ for 
$X$-particle with $m_X \geq 10^{13}$~GeV needs $S>44$. 
This value is within the 
range of the allowed values as discussed in Ref. \cite{KLLS}.

Let us now turn to the cosmological production of $X$-particles with 
$m_X \geq 10^{13}$~GeV. Several mechanisms were identified in \cite{BKV}, 
including 
thermal production at the reheating stage, production through the decay of 
inflaton field at the end of the "pre-heating"
period following inflation, and through the decay of hybrid topological 
defects, such as monopoles connected by strings or walls bounded by
strings.  

For the thermal production, temperatures comparable to $m_X$ are needed. 
In the case of a heavy decaying gravitino,
the reheating temperature $T_R$ (which is the highest temperature 
relevant for our problem)            
is severely limited to value below $10^8- 10^{10}$~GeV, depending 
on the gravitino mass (see Ref. \cite{ellis} and references therein).  
On the other hand, 
in models with dynamically broken supersymmetry, the lightest 
supersymmetric particle is the gravitino. Gravitinos with mass 
$m_{3/2} \leq 1$~keV  interact relatively strongly with the thermal bath,
thus decoupling relatively late, and can be the CDM particle \cite{grav}. 
In this scenario all phenomenological
constraints on $T_R$ (including the decay of the second 
lightest supersymmetric particle) disappear and one can assume
$T_R \sim 10^{11} - 10^{12}$~GeV. In this 
range of temperatures, $X$-particles are not in thermal equilibrium.
If $T_R < m_X$, the density  $n_X$ of $X$-particles produced during the 
reheating phase at time $t_R$ due to $a+\bar{a} \to X+\bar{X}$ is easily 
estimated as
\be
n_X(t_R) \sim N_a n_a^2 \sigma_X t_R \exp(-2m_X/T_R),
\label{eq:dens}
\ee 
where $N_a$ is the number of flavors which participate in the production of 
X-particles, $n_a$ is the density of $a$-particles and $\sigma_X$ is 
the production cross-section. The density of $X$-particles at the
present epoch can be found by the standard procedure of calculating
the ratio $n_X/s$, where 
$s$ is the entropy density. Then for $m_X = 1\cdot 10^{13}$~GeV
and $\xi_X$ in the wide range of values $10^{-8} - 10^{-4}$, the required
reheating temperature is $T_R \sim 3\cdot 10^{11}$~GeV.

In the second scenario mentioned above, non-equilibrium inflaton decay,
$X$-particles are usually overproduced and a second period of 
inflation is needed 
to suppress their density.

Finally, $X$-particles could be produced by TD such as strings or textures. 
Particle production occurs at string intersections or in collapsing texture 
knots. The evolution of defects is scale invariant, and roughly a constant 
number of particles $\nu$ is produced per horizon volume $t^3$ per Hubble 
time $t$. ($\nu \sim 1$ for textures and $\nu \gg 1$ for strings.) The main 
contribution to to the X-particle density is given by the earliest epoch,
soon after defect formation, and we find 
$\xi_X \sim 10^{-6} \nu (m_X/10^{13}~GeV)(T_f/10^{10}~GeV)^3$, where 
$T_f$ is the defect formation temperature. Defects of energy scale
$\eta \gtrsim m_X$ could be formed at a phase transition at or slightly 
before the end of inflation. In the former case, $T_f \sim T_R$ , while in 
the latter case defects should be considered as "formed" when their typical 
separation becomes smaller than $t$ (hence $T_f < T_R$). It should be noted 
that early evolution of defects may be affected by friction; our estimate 
of $\xi_X$ will then have to be modified. X particles can also be produced 
by hybrid topological defects: monopoles connected by strings or walls 
bound by strings. The required values of $n_X/s$ can be obtained for a wide 
range of defect parameters.

The decays of $X$-particles result in the production of nucleons with a 
spectrum $W_N(m_X,x)$, where $m_X$ is the mass of the X-particle 
and $x=E/m_X$. The flux of nucleons $(p,\bar{p},n,\bar{n})$ from the halo and 
extragalactic space can be calculated as
\be
I_N^{i}(E)={1\over{4\pi}}{n_X^i\over{\tau_X}}R_i{1\over{m_X}}W_N(m_X,x),
\label{eq:nhalo}
\ee
where index $i$ runs through $h$ (halo) and $ex$ (extragalactic), 
$R_i$ is the size of the halo $R_h$, or the attenuation length of 
UHE protons due to their collisions with microwave photons,      
$\lambda_p(E)$, for the halo case and extragalactic case, respectively. We 
shall assume $m_Xn_X^h=\xi_X\rho_{\rm CDM}^h$ and 
$m_Xn_X^{\rm ex}=\xi_X\Omega_{\rm CDM}\rho_{\rm cr}$, 
where $\xi_X$ describes the fraction 
of $X$-particles in CDM, $\Omega_{\rm CDM}$ is the CDM 
density in units of the critical density $\rho_{\rm cr}$, and 
$\rho_{CDM}^h \approx 0.3~GeV/cv^3$ is the CDM density in the halo.
We shall use the following values for these parameters: a large 
DM halo with $R_h=100$~kpc (a smaller halo with $R_h=50$~kpc is possible, 
too), $\Omega_{CDM}h^2=0.2$, 
the mass of $X$-particle in the range 
$10^{13}~{\rm GeV}<m_X<10^{16}$~GeV, 
the fraction of $X$-particles 
$\xi_X\ll 1$ and $\tau_X \gg t_0$, where $t_0$ is the age of the Universe.
The two last parameters are convolved in the flux calculations in a single
parameter $r_X=\xi_X t_0/\tau_X$. For $W_N(m_X,x)$
we shall use like in the previous section the QCD fragmentation function in 
MLLA approximation.
For the attenuation length of UHE protons due to their interactions with
microwave photons, we use the values given in the book \cite{BBDGP}. 

The high energy photon flux is produced mainly due to decays of neutral 
pions and can be calculated for the halo case as
\be
I^h_{\gamma}(E)=\frac{1}{4\pi}\frac{n_X}{\tau_X}R_h N_{\gamma}(E),
\label{gfl}
\ee
where  
$N_{\gamma}(E)$ is the number of photons with energy $E$ produced per
decay of one $X$-particle, which is given by Eq.(\ref{gnumber})

For the calculation of the extragalactic gamma-ray flux, it is enough to 
replace the size of the halo, $R_h$, by the absorption length of a photon, 
$\lambda_{\gamma}(E)$. The main photon absorption process is 
$e^+e^-$-production on background radiation and, at $E>1\cdot 10^{10}$~GeV,
on the radio background. The neutrino flux calculation is similar.

Before discussing the obtained results, we consider 
the astrophysical constraints.

The most stringent constraint comes from electromagnetic 
cascade radiation, discussed in the previous section.
In the present case 
this constraint is weaker, because the low-energy extragalactic
nucleon flux 
is $\sim 4$ times 
smaller than that one from the Galactic halo (see Fig.~2). Thus 
the cascade radiation is suppressed by the same factor. 

The cascade energy density calculated by integration over cosmological epochs
(with the dominant contribution given by the present epoch $z=0$) yields
in our case
\be
\omega_{\rm cas}=\frac{1}{5}r_X\frac{\Omega_{CDM}\rho_{cr}}{H_0t_0}=
6.3\cdot10^2 r_X f_{\pi}~{\rm eV/cm}^3.
\ee

To fit the UHE CR observational data by nucleons from halo, 
we need $r_X=5\cdot 10^{-11}$. Thus the cascade energy density is 
$\omega_{\rm cas}=3.2\cdot 10^{-8} f_{\pi}$~eV/cm$^3$, well below the
observational bound.

Let us now discuss the obtained results.
The fluxes shown in Fig.~2 are obtained for $R_h=100$~kpc, 
$m_X=1\cdot 10^{13}$~GeV and
$r_X=\xi_X t_0/\tau_X=5\cdot10^{-11}$. This ratio $r_X$ allows very small 
$\xi_X$ and $\tau_X > t_0$. The fluxes 
near the maximum energy $E_{\rm max}=5\cdot 10^{12}$~GeV  were only roughly
estimated (dotted lines on the graph). 

It is easy to verify that the extragalactic nucleon flux at $E \leq 
3\cdot 10^{9}$~GeV is suppressed by a factor $\sim 4$ and by a much larger 
factor at higher energies due to energy losses. The flux of 
extragalactic photons is suppressed even stronger, because the attenuation 
length for photons (due to absorption on radio-radiation) is much smaller
than for nucleons (see Ref. \cite{PB}). This flux is not shown in the graph.  
The flux of  high energy gamma-radiation from the halo is by a factor $7$ 
higher than that of nucleons and the neutrino flux, given in the Fig.2 as 
the sum of the dominant halo component and subdominant extragalactic one,
is twice higher than the gamma-ray flux.

The spectrum of the observed EAS is formed due to fluxes of gamma-rays and 
nucleons. The gamma-ray contribution to this spectrum is rather complicated.
In contrast to low energies, the photon-induced showers at 
$E>10^9$~GeV have the low-energy muon component as abundant as that 
for nucleon-induced showers \cite{AK}. However, the  
shower production by the photons is suppressed by the 
LPM effect \cite{LPM}
and by absorption in geomagnetic field (for recent calculations and 
discussion see \cite{ps,Kasa} and references therein). 

We wish to note that the excess of the gamma-ray flux over the nucleon
flux  from the halo is an unavoidable feature of this model. It follows
from the more effective production of pions 
than nucleons in the QCD cascades from the decay of $X$-particle. 

The signature of our model  might be the signal from the Virgo 
cluster. The virial mass of the Virgo cluster is
$M_{\rm Virgo} \sim 1\cdot 10^{15} M_{\odot}$ and the distance to it 
$R= 20$~Mpc. If UHE protons (and antiprotons) propagate rectilinearly from 
this source
(which could be the case for $E_p \sim 10^{11} - 10^{12}$~GeV), their 
flux is given by
\be
F_{p,\bar{p}}^{\rm Virgo}= r_X \frac{M_{\rm Virgo}}{t_0 R^2 m_X^2}W_N(m_X,x).
\ee    
The ratio of this flux to the diffuse flux from the half hemisphere is
$6.4\cdot 10^{-3}$. This signature becomes less pronounced at smaller 
energies, when protons can be strongly deflected by intergalactic magnetic 
fields.

\section{LSP IS UHE CARRIER}

LSP is the Lightest Supersymmetric Particle. It can be stable if R-parity 
is strictly conserved or unstable if R-parity is violated. To be able to 
reach the Earth from most remote regions in the Universe, the LSP must have 
lifetime longer than $\tau_{LSP} \gtrsim t_0/\Gamma$, where $t_0$ is the 
age of the Universe and $\Gamma=E/m_{LSP}$ is the Lorentz-factor of the LSP.
In case $m_{LSP} \sim 100~GeV$, $\tau_{LSP} > 1~yr$.

Theoretically the best motivated candidates for LSP are the neutralino and 
gravitino. We shall not consider the latter, because it is practically 
undetectable as UHE particle.

In all elaborated SUSY models the gluino is not the LSP. Only, if
the dimension-three SUSY breaking terms are set
to zero by hand, gluino with mass $m_{\tilde g}={\mathcal O}(1~GeV)$ can be the 
LSP \cite{fa96}. There is some controversy if
the low-mass window $1~GeV \lesssim m_{\tilde g} \lesssim 4~GeV$ for
the gluino is still allowed \cite{pdg,aleph}. Nevertheless, we shall study
the production of high-energy gluinos and their interaction with matter
being inspired by the recent suggestion \cite{Farr} (see also \cite{MN}), 
that the atmospheric showers observed at the highest energies can be
produced by colorless hadrons containing gluinos. 
We shall refer to any of such hadron as
$\tilde{g}$-hadron. Light gluinos as 
UHE particles with energy $E \gtrsim 10^{16}$~eV were  considered in 
some detail in the literature in connection with Cyg X-3 \cite{aur,BI}.
Additionally, we consider heavy gluinos  with $m_{\tilde g} \gtrsim
150$~GeV \cite{MN}.

UHE LSP are most naturally produced at the decays of unstable superheavy 
particles, either from TD or as the relic ones \cite{BK}. 

The QCD parton cascade is not a unique cascade process. A cascade 
multiplication of partons at the decay of superheavy particle appears 
whenever a probability of production of extra parton has the terms 
$\alpha \ln Q^2$ or $\alpha \ln^2 Q^2$, where $Q$ is a maximum of parton 
transverse momentum, i.e. $m_X$ in our case. Regardless of smallness of 
$\alpha$, the cascade develops as long as
 $\alpha \ln Q^2 \gtrsim 1$. Therefore, for 
extremely large $Q^2$ we are interested in, a cascade develops due to
parton multiplication through $SU(2)\times U(1)$ interactions as well. Like in 
QCD, the account of diagrams with $\alpha \ln Q^2$ gives the 
Leading Logarithm Approximation to the cascade fragmentation function.

For each next generation of cascade particles the virtuality of partons 
$q^2$ diminishes. When $q^2 \gg m_{SUSY}^2$ , where $m_{SUSY}$ is a typical 
mass of supersymmetric particles, the number of supersymmetric partons 
in the cascade is the same as their ordinary partners. At $q^2 < 
m_{SUSY}^2$  the supersymmetric particles are not produced any more and 
the remaining particles decay producing the LSP.  In Ref.(\cite{BK}) 
a simple Monte Carlo simulation for SUSU cascading was performed 
and the spectrum of emitted LSP was calculated. LSP take away a 
considerable fraction of the total energy ($\sim 40\%$).  

The fluxes of UHE LSP are shown in Fig.~3 for the case of their production 
in cosmic necklaces (see section II). When the LSP  is neutralino, the 
flux is somewhat lower than neutrino flux. The neutralino-nucleon cross-
section, $\sigma_{\chi N}$, is also smaller than that for neutrino. For the 
theoretically favorable masses of supersymmetric particles, 
$\sigma_{\chi N} \sim 10^{-34}~cm^2$ at extremely high energies. If the the 
masses of squarks are near their experimental bound, $M_{L,R} \sim 180~GeV$, 
the cross-section is 60 times higher.

{\em Gluino as the LSP} is another phenomenological option. Let us discuss 
shortly the status of the gluino as LSP.

In all elaborated SUSY models the gluino is not LSP, and this possibility 
is considered on purely phenomenological basis.
 Accelerator experiments give the lower limit on the gluino mass as 
$m_{\tilde{g}} \gtrsim 150$~GeV \cite{pdg}. The upper limit of the 
gluino mass is given by cosmological and astrophysical constraints, as 
was recently discussed in \cite{MN}. In this work it 
was shown that if the gluino provides the dark matter observed in our galaxy,
the signal from gluino annihilation and the abundance of anomalous heavy 
nuclei is too high. Since we are not interested in the case when gluino 
is DM particle, we can use these arguments to obtain an upper limit for 
the gluino mass. Calculating the relic density of 
gluinos (similar as in \cite{MN}) and using
the condition $\Omega_{\tilde{g}} \ll \Omega_{\rm CDM}$, we obtained 
$m_{\tilde{g}} \ll 9$~TeV.

Now we come to very interesting argument against existence  
of a light stable or quasistable gluino \cite{VO}.
It is plausible that the {\em glueballino} ($\tilde{g}g$) is the lightest 
hadronic 
state of gluino \cite{aur,BI}. However, {\em gluebarino\/}, i.e. the bound 
state of gluino and three quarks, is almost stable because baryon number 
is extremely weakly violated. In Ref.~\cite{VO} it is argued that the 
lightest gluebarino is the neutral state ($\tilde{g}uud$).  
 These charged gluebarinos are produced by cosmic
rays in the earth atmosphere \cite{VO}, and light gluino as LSP is 
excluded by the search for heavy hydrogen or by proton decay 
experiments (in case of quasistable gluino). In the case that the lightest 
gluebarino is neutral, see \cite{fa96}, the arguments of \cite{VO}
still work if a  neutral gluebarino forms a bound state with the nuclei.  
Thus, a light gluino is disfavored.

The situation is different if the gluino is heavy, 
$m_{\tilde{g}}\gtrsim 150~GeV$.  This gluino can be unstable
due to weak R-parity violation \cite{BJV} and have a lifetime 
$\tau_{\tilde{g}} \gtrsim 1$~yr, {\it i.e.\/}
long enough to be UHE carrier (see beginning of this section).  
Then the calculated relic density at the time 
of decay is not in conflict with the cascade nucleosynthesis and all 
cosmologically produced $\tilde{g}$-hadrons decayed up to the present time.
Moreover, the production of these gluinos by cosmic rays in the
atmosphere is ineffective because of their large mass.

Glueballino, or more generally $\tilde{g}$-hadron, looses its energy while 
propagating from a source to the Earth. The dominant energy loss of 
the $\tilde{g}$-hadron is due to pion production in collisions with microwave 
photons. Pion production effectively starts at the same Lorentz-factor as 
in the case of the proton. This implies that the energy of the GZK 
cutoff is a factor $m_{\tilde{g}}/m_p$ higher than in case of the proton. 
The attenuation length also increases because the fraction of energy lost 
near the threshold is small, $\mu/m_{\tilde{g}}$, where $\mu$ is a pion 
mass. Therefore, even for light $\tilde{g}$-hadrons, $m_{\tilde{g}} \gtrsim 
2~GeV$, the steepening of the spectrum is less pronounced than for 
protons.    

The spectrum of $\tilde{g}$-hadrons from the cosmic necklaces accounted for 
absorption in intergalactic space, is shown in Fig.~3.

A very light UHE $\tilde{g}$-hadron interacts with the nucleons in the 
atmosphere similarly to UHE proton. The cross-section is reduced 
only due to the radius of $\tilde{g}$-hadron and is of order of 
$\sim 1~mb$ \cite{BI}. In case of very heavy $\tilde{g}$-hadron the total 
cross-section can be of the same order of magnitude, but the 
cross-section with the large energy transfer, relevant for the 
detection in the atmosphere, is very small \cite{BK}. This is due to the 
fact that interaction of gluino in case of large energy transfer is 
characterized by large $Q^2$ and thus interaction is a deep inelastic 
QCD scattering.  

Thus, only UHE gluino from low-mass window 
$1~GeV \leq m_{\tilde{g}}\leq 4~GeV$ could be a candidate for observed 
UHE particles, but it is disfavored by the arguments given above. 

\section{CONCLUSIONS}

Topological Defects naturally produce particles with extremely high 
energies, much in excess of what is presently observed. However, the fluxes 
from most known TD are too small. So far only necklaces \cite{BV} and 
monopole-antimonopole pairs \cite{BS} can provide the observed flux of UHE CR. 

Another promising sources of UHE CR are relic superheavy particles 
\cite{KR,BKV}. These particles should be clustering in the halo of 
our Galaxy \cite{BKV}, and thus UHE CR produced at their decays do not 
have the GZK cutoff. The signatures of this model are dominance of 
photons in the primary flux and Virgo cluster as a possible discrete source.

Apart from protons, photons and neutrinos the UHE carriers can be 
neutralinos \cite{BK}, gluino \cite{Farr,MN,BK} and monopoles \cite{Weil,MN}.
While neutralino is a natural candidate for the Lightest Supersymmetric 
Particle (LSP) in SUSY models, gluino can be considered as LSP only 
phenomenologically. LSP are naturally produced in the parton cascade at 
the decay of superheavy X-particles. In case of neutralino both fluxes and 
cross-sections for interaction is somewhat lower than for neutrino. In case 
of gluinos the fluxes are comparable with that of neutralinos, but 
cross-sections for the production of observed extensive air showers are 
large enough only for light gluinos. These are disfavored, especially if 
the charged gluebarino is lighter than the neutral one \cite{VO}. \\*[1mm]
\begin{center}
ACKNOWLEDGEMENTS
\end{center}
\vspace{1mm}
This report is based on my recent works with Michael Kachelriess and 
Alex Vilenkin \cite{BV,BKV,BK}. I am grateful to my co-authors for 
pleasant and useful cooperation and for many discussions.  

Many thanks are to the organizers of the workshop for the most efficient 
work. I am especially grateful to Jonathan Ormes for all his efforts as the 
Chairman of the Organizing Committee and for inviting me to this most 
interesting meeting.


\phantom{}
\begin{figure}[t]
\vspace{19.0cm}
\includegraphics{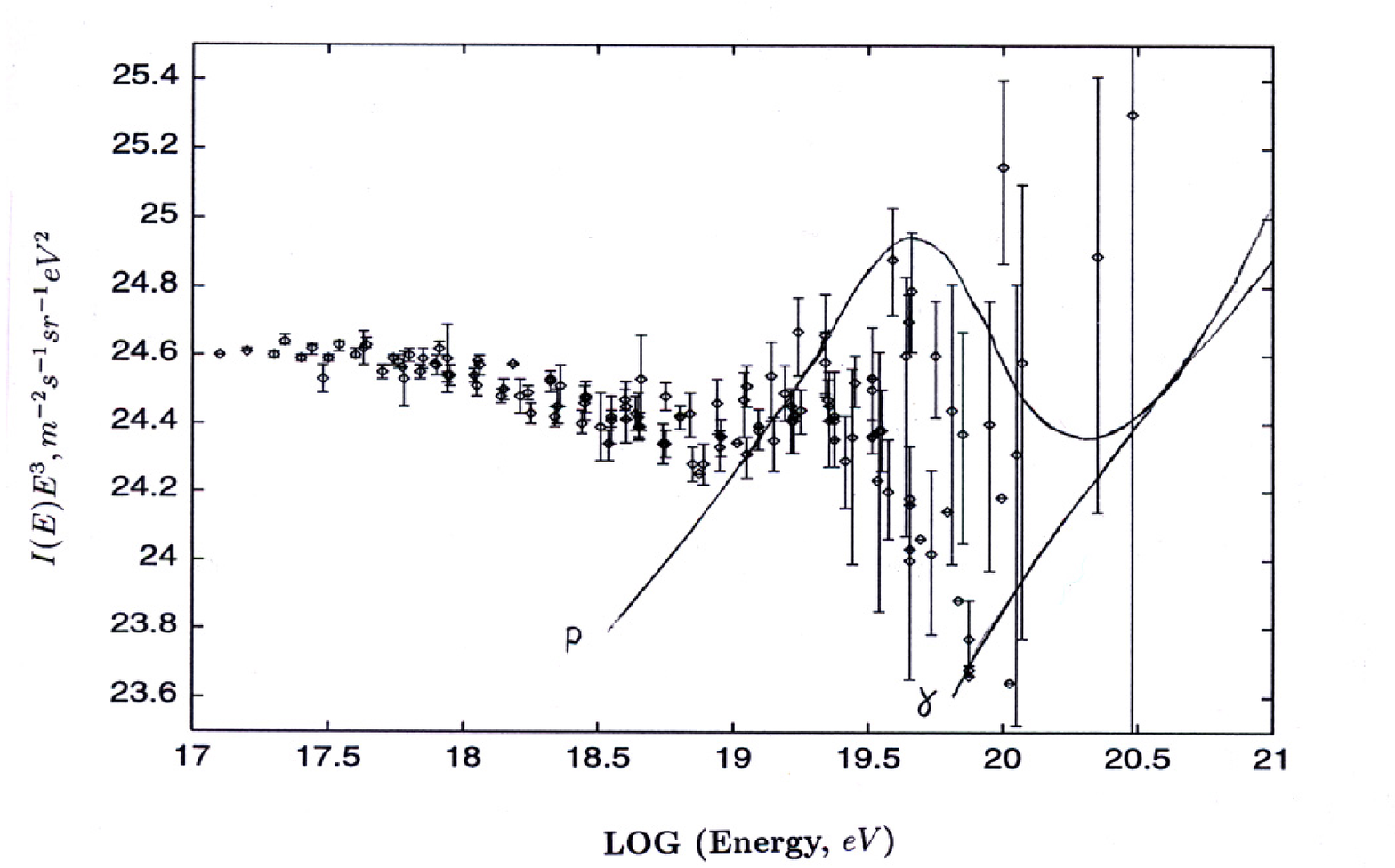}
\caption{\label{fig1}
Predicted proton (p) and gamma-ray ($\gamma$) fluxes from necklaces
in comparison with experimental data.}
\end{figure}

\phantom{}
\begin{figure}[t]
\vspace{19.0cm}
\includegraphics{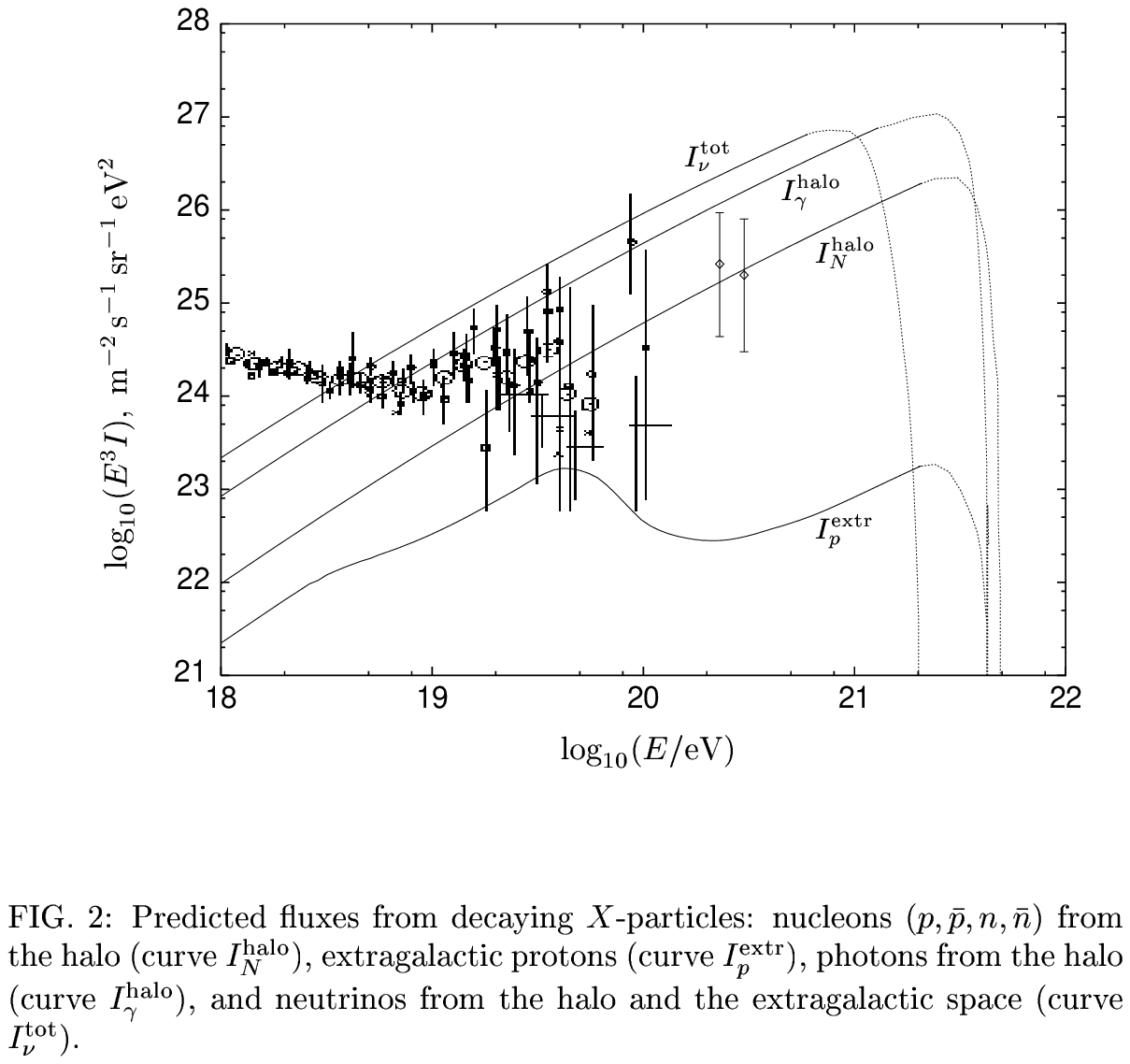}
\end{figure}

\phantom{}
\begin{figure}[t]
\vspace{19.0cm}
\includegraphics{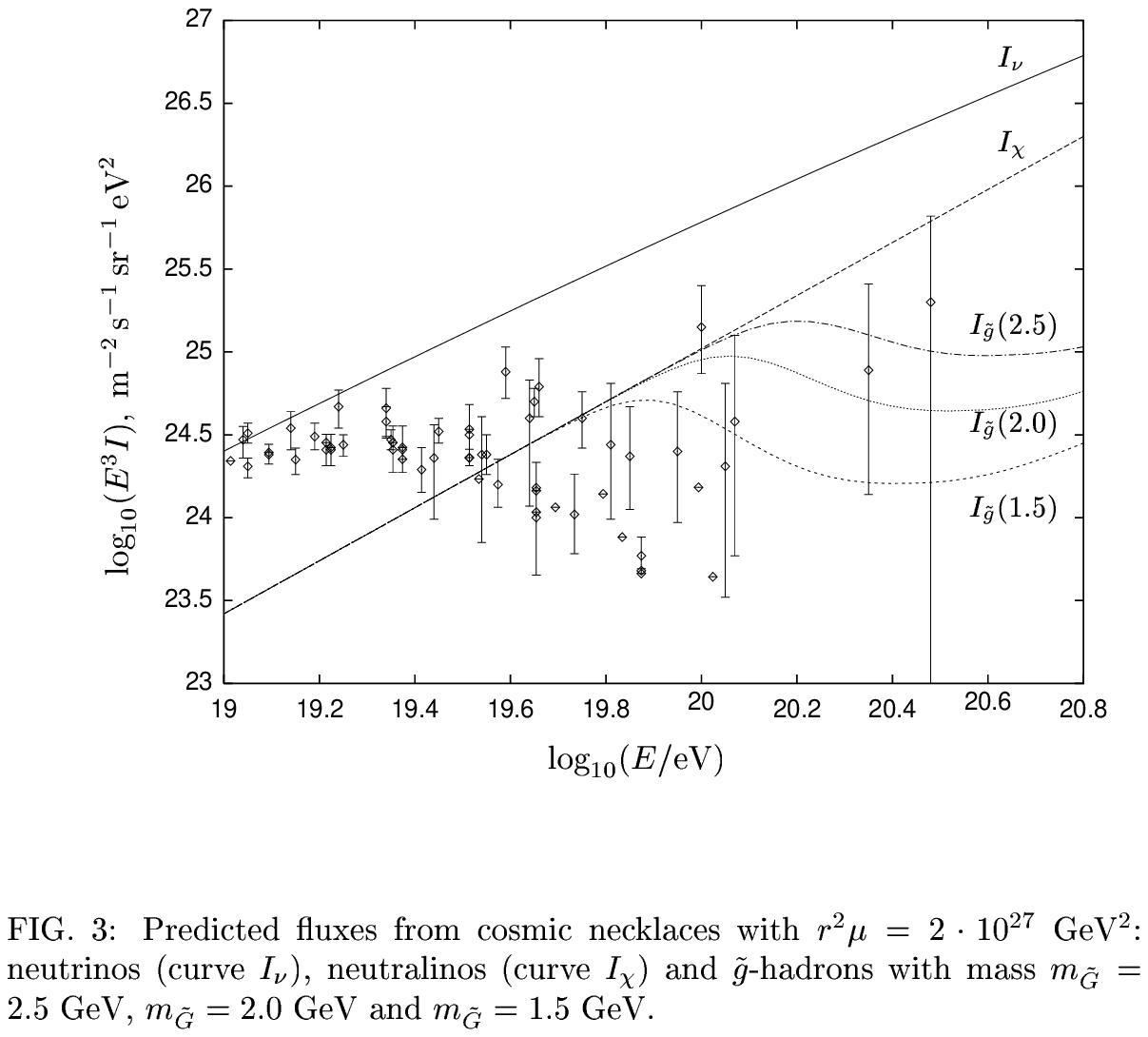}
\end{figure}

\end{document}